\documentclass[letterpaper,twocolumn,10pt]{article}
\usepackage{usenix2022_SOUPS}

\usepackage{tikz}
\usepackage{amsmath}

\usepackage{algorithmic}
\usepackage{graphicx}
\usepackage{textcomp}
\usepackage{bmpsize}
\usepackage{xcolor}
\usepackage{lipsum}
\usepackage{comment}
\usepackage{tikz}
\usepackage{amsmath}
\usepackage{todonotes}
\usepackage{enumitem}
\usepackage{verbatim}
\usepackage{multirow}
\usepackage{subcaption}

\newcommand*\samethanks[1][\value{footnote}]{\footnotemark[#1]}

\begin{document}

\date{}

\title{\Large \bf Buying Privacy: User Perceptions of Privacy Threats from Mobile Apps}

\def\plainauthor{Author name(s) for PDF metadata. Don't forget to anonymize for submission!}

\author{
\hspace{.2in}{\rm Jenny Tang\thanks{Work was done while Tang and Lerner were at Wellesley College.}}\\
\hspace{.2in}Carnegie Mellon University
\and
{\rm Hannah Shoemaker}\\
Pomona College 
\and
\hspace{0in}{\rm Leah Teffera}\hspace{.4in}\\
\hspace{.2in}Wellesley College\hspace{.2in}
\and
{\rm Eleanor Birrell} \\
Pomona College
\and
{\rm Ada Lerner\samethanks} \\
Northeastern University
} 

\maketitle

\begin{abstract}
As technology and technology companies have grown in power, ubiquity, and societal influence, some companies---and notably some mobile apps---have come to be perceived as privacy threats. Prior work has considered how various factors impact perceptions of threat, including social factors, political speech, and user-interface design. In this work, we investigate how user-visible context clues impact perceptions about whether a mobile application application poses a privacy threat.  We conduct a user study with 2109 users in which we find that users depend on context clues---such as presence of advertising and occurrence (and timing of payment)---to determine the extent to which a mobile app poses a privacy threat. We also quantify how accurately user assessments match published data collection practices, and we identify a commonly-held misconception about how payments are processed. This work provides new insight into how users assess the privacy threat posed by mobile apps and into social norms around data collection.
\end{abstract}

\section{Introduction}


Although there is much work identifying users' mental models of threats and individuals' security and privacy perceptions and behaviors (e.g.,~\cite{malhotra_internet_2004,nissenbaum2004privacy,egelman2015scaling,Wash2010FolkMO,kang2015,Lin2012ExpectationAP}),
there has been little examination of what factors influence which principals come to be perceived as security or privacy threats. This work bridges that gap in the context of mobile game apps. 
We ask: what user-visible signals influence perceptions of privacy threats? We consider two possible signals that have been found to be correlated with actual data use practices: advertising~\cite{stevens2012,book2013longitudinal} and price (free vs. paid)~\cite{saborido2017comprehension,han2020,bamberger2020}. Prior work has found that these signals in combination affect user perceptions of apps (e.g., users expect that a paid, ad-free version of the app will have better security and privacy behaviors than a free, ad-supported app~\cite{han2020,bamberger2020}); this work takes the next step by exploring how each of these signals individually impact perceptions about the privacy threat posed by mobile apps. We also consider a third possible signal---timing of payment---that is not known to correlate with actual data use practices, but for which norms on the major app stores have changed over time (away from paying for an ad-free version of an app and towards making an in-app purchase to remove ads from an ad-supported free app); our initial hypothesis was that users would not interpret the timing of payment as a signal about the degree of privacy threat posed by an app. 

We conducted an online user survey ($n=2109$) to measure whether these signals persuade users that a mobile app should be perceived as a \emph{privacy threat}, which we characterize as an principal that is perceived as both (1) untrustworthy and (2) having access to personal information. Thus our studies measured changes in perceived trustworthiness and perceived data collection by apps in order to operationalize perceptions of apps as privacy threats.


Our first research question is: how do each of our signals (presence of ads, occurrence of payment, timing of payment) individually affect user perceptions of privacy threats? We found that respondents who were assigned to the conditions with ads assessed apps as posing greater privacy threats.  We found that  presence or absence of payment alone did not have a significant impact on users' assessment. We also found that users were not able to accurately assess which data types were collected by a particular app. While conducting this analysis, we also observed that  many users hold misconceptions about what data is collected (and by whom) for the purpose of payment processing. 


However, we found evidence to disprove our timing hypothesis. We found that users assessed apps that initially contained ads as posing a significantly greater privacy threat---both collecting more data and being less trustworthy---after paying to remove ads from a previously installed app compared to an (ad-free) app they had initially paid for. 


We view the key contributions of this work as:
\begin{enumerate}
    \item Identifying and experimentally validating three factors---advertising, payment, and timing of payment---that influence perceptions of the privacy threat posed by mobile game apps.
    \item Establishing that people use ads as a proxy for assessing privacy threat in the context of mobile game apps.
    \item Quantifying social norms around data collection by mobile apps and the intersection between those norms and user-visible signals.
    \item Identifying and quantifying misconceptions about how data is collected in order to process payments.
\end{enumerate}

Further work will be required to determine whether there exist other factors that influence user assessment and to extend the scope beyond game apps. Nonetheless, we view this work as an important step in providing a more nuanced understanding of how users evaluate the privacy threat posed by mobile apps and of current social norms about data collection by these apps.

\section{Related Work}
\label{sec:relatedWork}

To the best of our knowledge, this is the first work to specifically examine the effects of advertising, payment, and timing of payment on user assessments of privacy threats from mobile game apps. However, the effect of other factors on the formation of attitudes towards technological threats and the relationships between advertising, payment, data collection, and trust have been investigated in a variety of other contexts. 

\subsection{Persuasion, Framing, and the Formation of Attitudes}

Existing multi-disciplinary literature has studied the various factors that influence the persuasiveness of claims and narratives regarding credibility and threat.  The elaboration likelihood model (ELM)~\cite{petty_elaboration_1986} takes into account both characteristics of the recipient of the argument as well as the context of the presented arguments, and has been widely-used in various disciplines to explain attitude change and formation. Within the security and privacy realm, it has been used to model the way consumers judge online review credibility, Android users' concerns and download intentions, and the effect of GDPR consent pop-ups~\cite{cheung2012review, gu_privacy_2017, cara2021gdpr}. 

A myriad of factors have been shown to influence user trust, such as the design of interfaces, reputation, social factors, expert opinions, as well as individual predispositions and preferences~\cite{kang2015, bart_are_2005, chen_influence_2015, parker_effects_2020, metzger_effects_2006, koufaris_development_2004, mcknight2002impact, olsina_mobileapp_2018, fogg_what_2001, yan_research_2011, kehr2013rethinking, fagan2016they, bart_are_2005, redmiles_i_2016}.
The quality of websites and interface design, can be strong predictors of user trust, with amateurish designs decreasing credibility and feeling `sketchy'' as indicated through visual cues~\cite{olsina_mobileapp_2018, mcknight2002impact,fogg_what_2001}. However, some of the influence of design may vary from individual to individual ~\cite{murayama2007structure, yan_research_2011}. Overall, personal traits and indvidual differences in disposition or assessment of risks and benefits regarding privacy tend to at least somewhat influence trust perceptions and security-related behaviors~\cite{malhotra_internet_2004, kehr2013rethinking, fagan2016they}, though they may not always have significant effects \cite{koufaris_development_2004}.
Another factor is reputation: better-known brands and apps garner more trust from consumers~\cite{kang2015, bart_are_2005, chen_influence_2015, parker_effects_2020, metzger_effects_2006, koufaris_development_2004}. Though other factors may influence reputation, most studies that investigate reputational benefits use popularity as a proxy for reputation~\cite{mcknight2002impact, metzger_effects_2006}.
Social factors including family, friends, and media are also influential in user trust and behavior~\cite{bart_are_2005, redmiles_i_2016, das2014effect, alohali2017information, mendel_susceptibility_2017}. Other environmental factors such as the context in which data collection occurs are also relevant to trust assessment and formation~\cite{nissenbaum2004privacy, camp_designing_2003, riegelsberger_mechanics_2005, mesch2012, kehr2013rethinking, Lin2012ExpectationAP}.  For example, users assessed data collection for financial services sites and travel sites differently given the differing contexts under which data collection was occurring~\cite{bart_are_2005}. Prior work has also found that users expect game apps to only collect data that is contextually appropriate for such an app~\cite{shklovski2014leakiness}, an expectation that does not always reflect actual data collection by mobile apps. We extend the prior work by exploring how user-visible signals---such as advertising, payment, and timing of payment---influence user trust.

\subsection{The Effect of Advertising}

Van Kleek et al.~\cite{van2018x} found that apps that displayed ads prominently were expected to sharing data with advertisers, and that those  those that did not were viewed as less likely to do so. Our work extends those findings to explore how advertising impacts assumptions about the amount of data collected by mobile apps as well as the apps' trustworthiness. Wang et al.~\cite{wang2015investigating} found that enhanced awareness of how personal data is used for advertising purposes significantly impacts users' perceptions of an app, their trust levels, and their actual behaviour. However, a study of user reviews found that the main costs users associated with ads were memory/CPU overhead, battery consumption, traffic usage, and number of ads, with the first two costs being the ones users were most concerned about rather than the less tangible costs of possible privacy loss~\cite{Gao2016IntelliAdUI}. Our work is distinct in that it studies how the mere presence or absence of advertising affects user conceptions of trust and data collection.

Most of the literature about advertising in the mobile ecosystem focuses on the actual data collection by ad libraries rather than perceptions of data collection. 
Several independent projects have found that most ad libraries collect private information about users~\cite{grace2012unsafe,Book_Bronk_2016,hu2019}, including users' call logs~\cite{grace2012unsafe}, precise location data~\cite{hu2019}, and various personally-identifiable information~\cite{Book_Bronk_2016}. 
The permissions requested by ad libraries have expanded over the years~\cite{book2013longitudinal,book2015} as has data collection through other channels~\cite{demetriou2016}. 
Ad libraries collect sensitive information and check for dangerous permissions without documenting this data collection~\cite{seo2016, stevens2012}.
Studies have concluded that in-app advertising and ad libraries have the potential to leak a wide range of user data~\cite{bookwallach2013, meng2016price}, and that ad libraries pose a security risk because some run code retrieved directly from the internet~\cite{grace2012unsafe} and because apps are slow to adopt new, patched versions~\cite{backes2016}. Finally, a line of work has demonstrated that malicious actors including libraries and advertisers can infer sensitive information about users through methods such as performing inter-library collusion attacks, accessing external storage, or simply purchasing advertisements~\cite{taylor2017intralibrary, son2016, Vines2017}.

\begin{figure*}[t!]
    \centering
    \begin{tabular}{lll}
    \cline{1-1}\cline{3-3}
    \textbf{Types of Data Collected} & \verb|      |& \textbf{Purposes of Data Collection}  \\
    \cline{1-1}\cline{3-3}
    phone number & & to sell to other companies \\
    email & &to make more money from ads  \\ 
    phone account information  & & to identify who uses their app  \\
    physical mailing address & & to improve the app  \\
     credit card information && to process payments   \\
     information that uniquely identifies the user  && to perform necessary app functionality \\
     phone identifier & & \\
     general information about the user's phone & & \\ 
     current IP address& & \\ 
     
     basic demographics & & \\ 
     
     sensitive demographics & & \\ 
     current location & &  \\ 
     interactions with the app & & \\ 
     interactions with ads in the app & & \\ 
     interactions with other apps& &  \\ 
     sensor data from the camera and/or microphone& & \\ 
     
     photos& & \\ 
     contact list& & \\ 
     
     information about the user's interests& & \\ 
     \cline{1-1}\cline{3-3}
    \end{tabular}
    \caption{Results of qualitatively coding responses from pilot study participants. These 19 data types and 6 purposes we offered as options for the multiple-choice questions about data collection included in our full study.}
    \label{fig:pilotresults}
\end{figure*}

\subsection{The Effect of Payment}

A recent line of work has compared users' security and privacy expectations for free versus paid versions of an app. Han et al.~\cite{han2019,han2020,bamberger2020} compared the data use practices of pairs of free Android apps and their paid counterparts. They also conducted an online user survey to explore users' expectations for paid apps. They found that users expected paid apps to be ad-free (a finding that might explain our result that users trusted the PaidWithAds app significantly less than the apps in other conditions). They also identified significant differences in expectations about data use practices---e.g., paid apps are considered less likely to share data with advertisers and paid apps are considered more likely to securely store data to protect it from potential breaches. This work extends their findings about user expectations by focusing on the impact of payment  on the \emph{amount} of data collection rather than on expectations about data use practices. We also consider two additional signals---presence of ads and timing of payment---independently, and we  evaluation the impact of our signals on an app's overall perceived trustworthiness.

Van Kleek et al.~\cite{van2018x} conducted a qualitative user study that found that free versions of apps were expected to send data to more companies than their paid counterparts because paid apps were perceived to need less ad support. Our work extends this with a large-scale study quantifying the impact of payment on assumptions about data collection. 

In addition to work about the impact of payment on user perceptions, there is prior work that has empirically measured differences in privacy and security between free and paid apps. Many of these studies have found that that free apps are less secure and offer less privacy than paid apps.  Seneviratne et al.~\cite{seneviratne2015} found that approximately 60\% of paid apps studied were connected to trackers that collected users’ personal information, while 85-95\% of free apps studied were connected to these trackers. Previous studies have also found that free apps typically request more permissions than paid apps~\cite{harrischin2016, chia2012, leontiadis2012, bamberger2020, saborido2017comprehension}. Kummer and Schulte~\cite{kummerschulte2019} found that the cheaper an app is, the more privacy-sensitive permissions it requests. Appthority~\cite{appthority2014} found that free apps are generally riskier than paid apps, with 66\% of free apps tracking for location, while only 37\% of paid apps perform this tracking. 
However, empirical studies have also shown that the privacy and security benefits of paid apps may be limited. Several studies have found that for more than 40\% of apps with both a paid and a free version available, the paid version contains all of the same third-party libraries as the free version~\cite{han2020, bamberger2020, han2019}. Watanabe et al.~\cite{watanabe2017} affirmed the danger of these libraries, finding that 70\% (resp. 50\%) of the vulnerabilities of free (resp. paid) apps, come from these software libraries. Studies have also found that more than 50\% of paid apps request all of the same dangerous permissions as their free counterparts~\cite{han2020, bamberger2020, han2019}. Finally, Appthority~\cite{appthority2014} found that while 99\% of free apps studied exhibited at least one dangerous behavior, 83\% of paid apps also exhibited at least one dangerous behavior.

\section{Methodology}
\label{sec:methods}

In the course of this work, we conducted two qualitative pilot studies and one full user study. The pilot studies asked open-ended questions about data collection, and the full study experimentally validated factors that influence user \emph{assessments} of data collection (i.e., the number of data types believed to be collected) and trust in apps (i.e., how trustworthy they rated the app). All studies were conducted on Amazon Mechanical Turk, and participation was restricted to workers located in the United States with at least 50 HITs and at least a 95\% accept rate. The full study included a open-response question that we used as an attention check question; participants who successfully completed the attention check question were compensated based on an estimated survey completion time at a prorated rate of \$12/hour. Participants were informed about our data storage and data use practices in advance, and no personally-identifiable information was collected.

\begin{figure*}[t!]
\begin{center}
\begin{subfigure}{.4\textwidth}
\includegraphics[width=\textwidth]{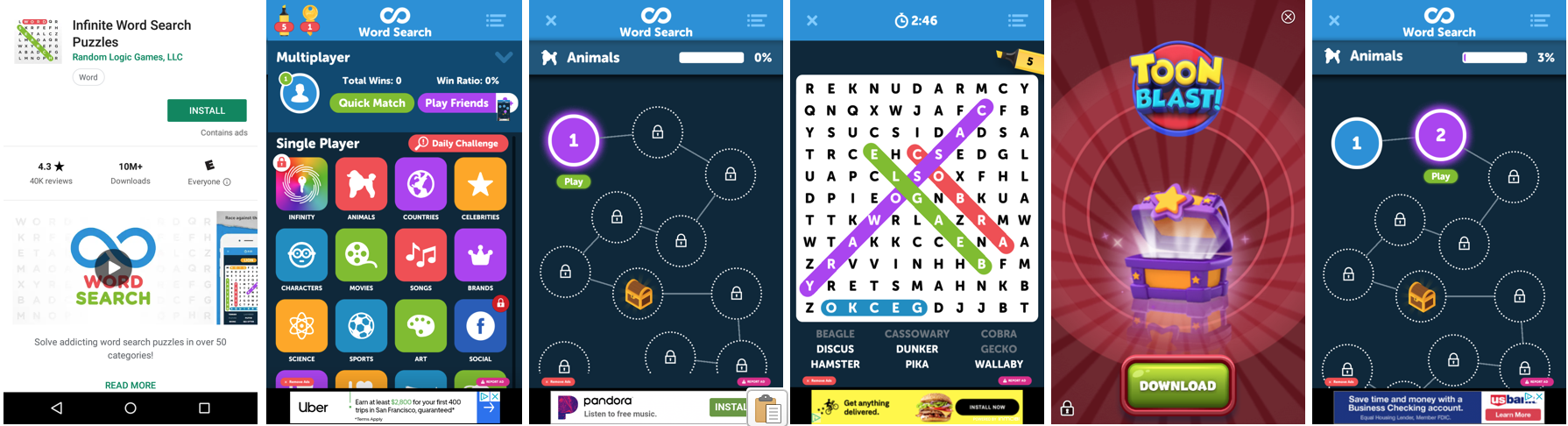}
\caption{FreeWithAds}\label{subfig:woody-FreeWithAds}
\end{subfigure}
\hspace{.04\textwidth}
\begin{subfigure}{.4\textwidth}
\includegraphics[width=\textwidth]{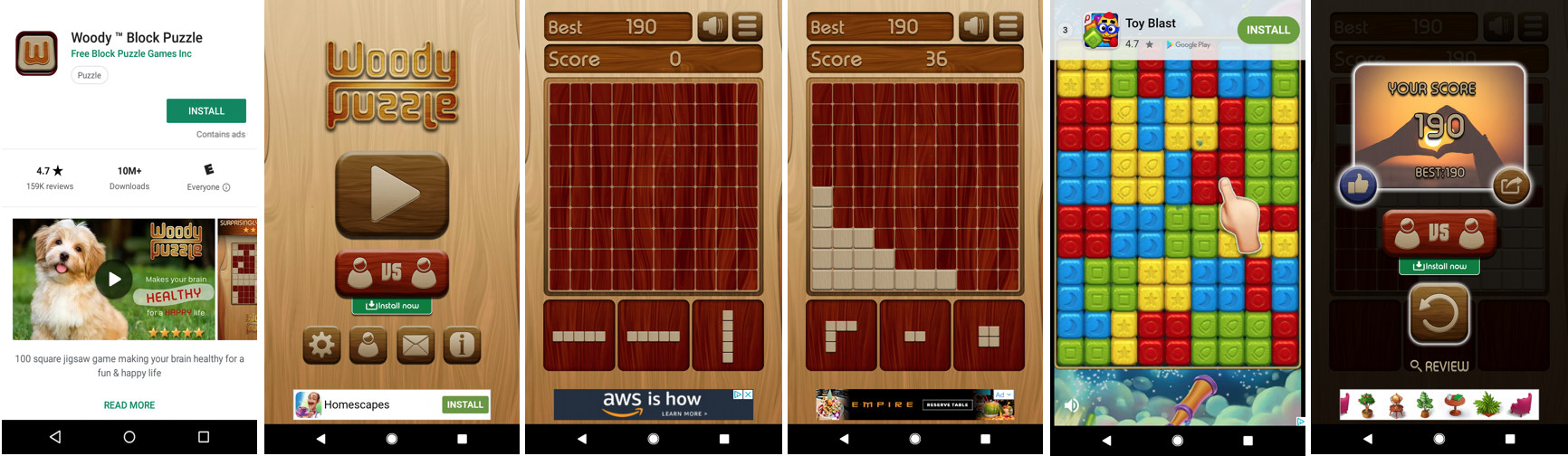}
\caption{FreeWithAds}\label{subfig:woody-FreeWithAds}
\end{subfigure}
\hfill
\begin{subfigure}{.4\textwidth}
\includegraphics[width=\textwidth]{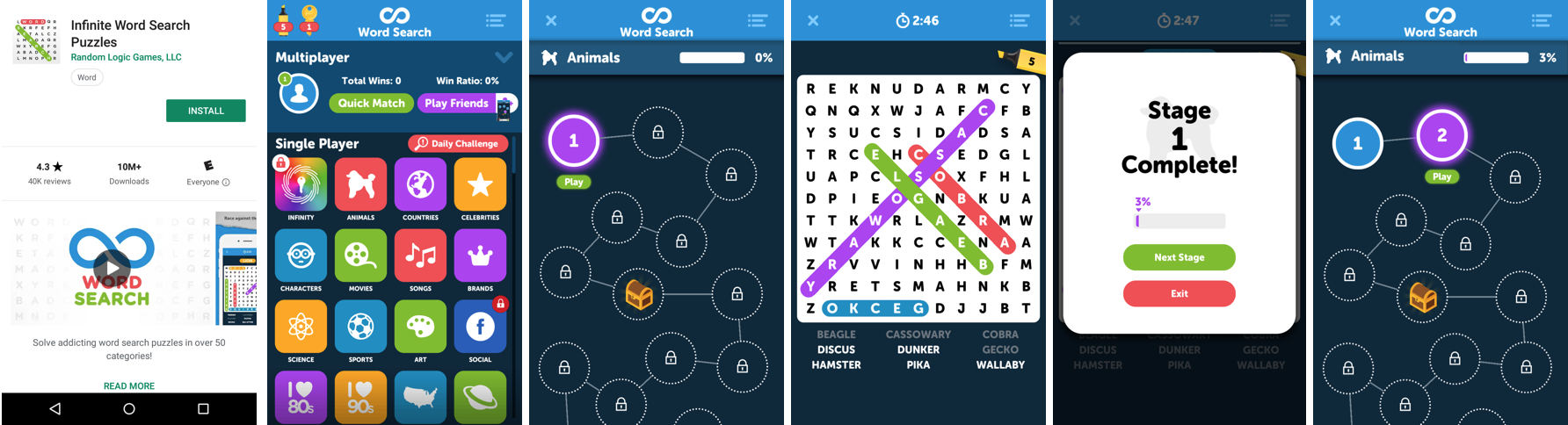}
\caption{FreeNoAds}\label{subfig:woody-FreeNoAds}
\end{subfigure}
\hspace{.04\textwidth}
\begin{subfigure}{.4\textwidth}
\includegraphics[width=\textwidth]{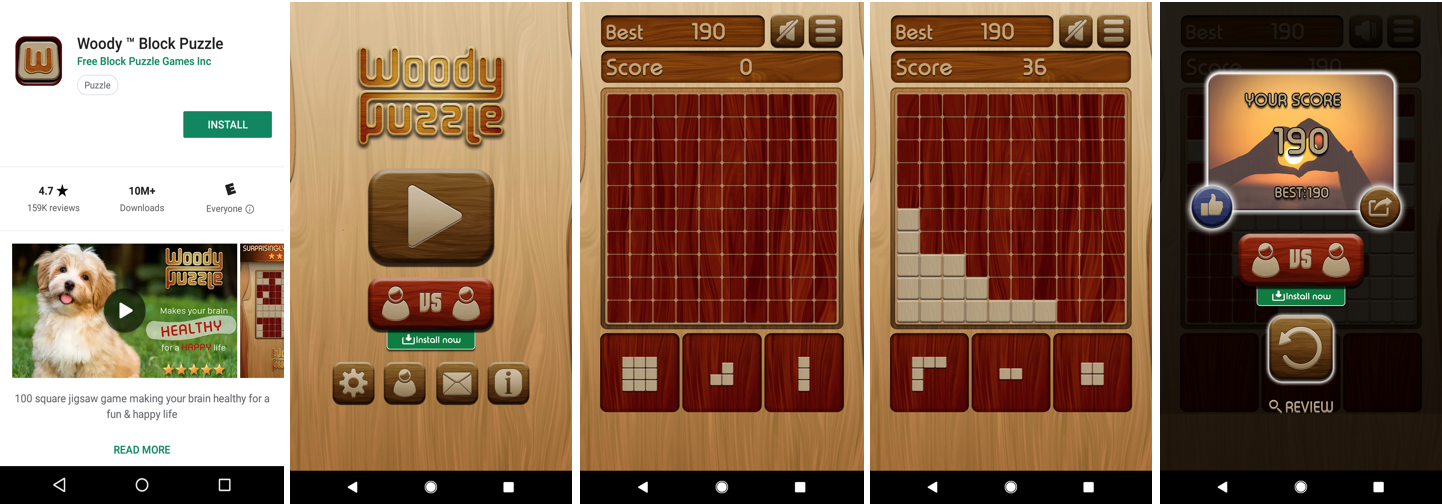}
\caption{FreeNoAds}\label{subfig:woody-FreeNoAds}
\end{subfigure}

\begin{subfigure}{.4\textwidth}
\includegraphics[width=\textwidth]{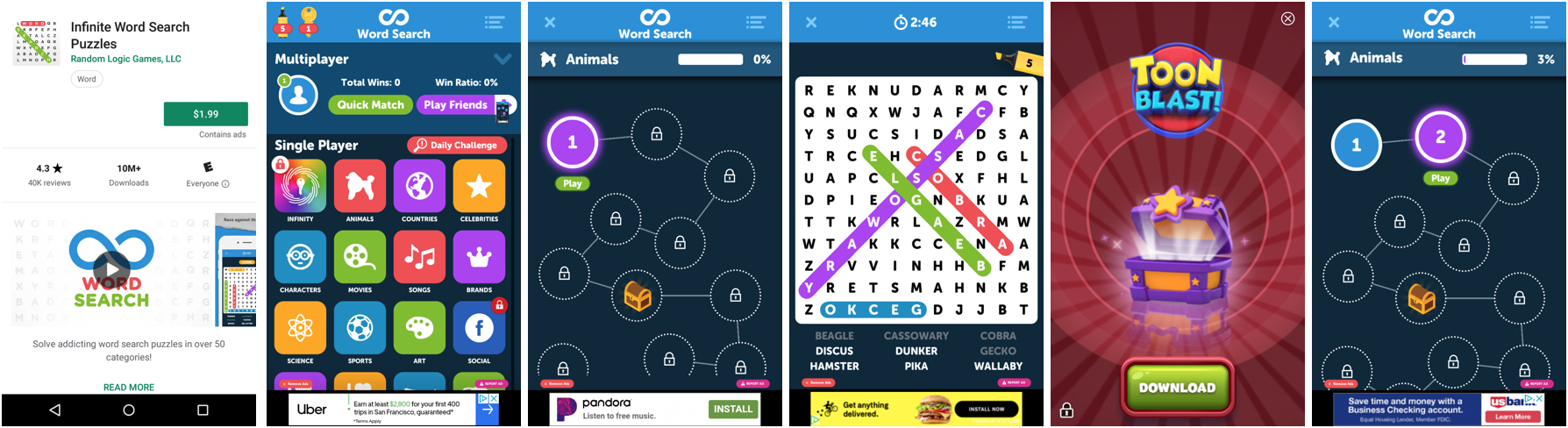}
\caption{PaidWithAds}\label{subfig:woody-PaidNoAds}
\end{subfigure}
\hspace{.04\textwidth}
\begin{subfigure}{.4\textwidth}
\includegraphics[width=\textwidth]{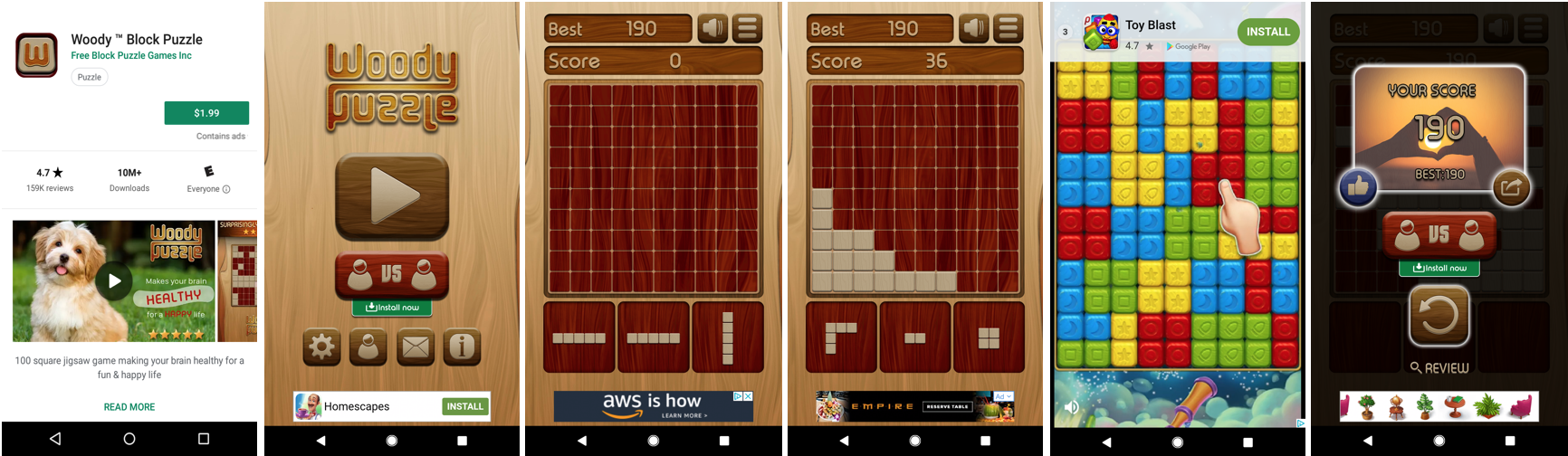}
\caption{PaidWithAds}\label{subfig:woody-PaidNoAds}
\end{subfigure}

\begin{subfigure}{.4\textwidth}
\includegraphics[width=\textwidth]{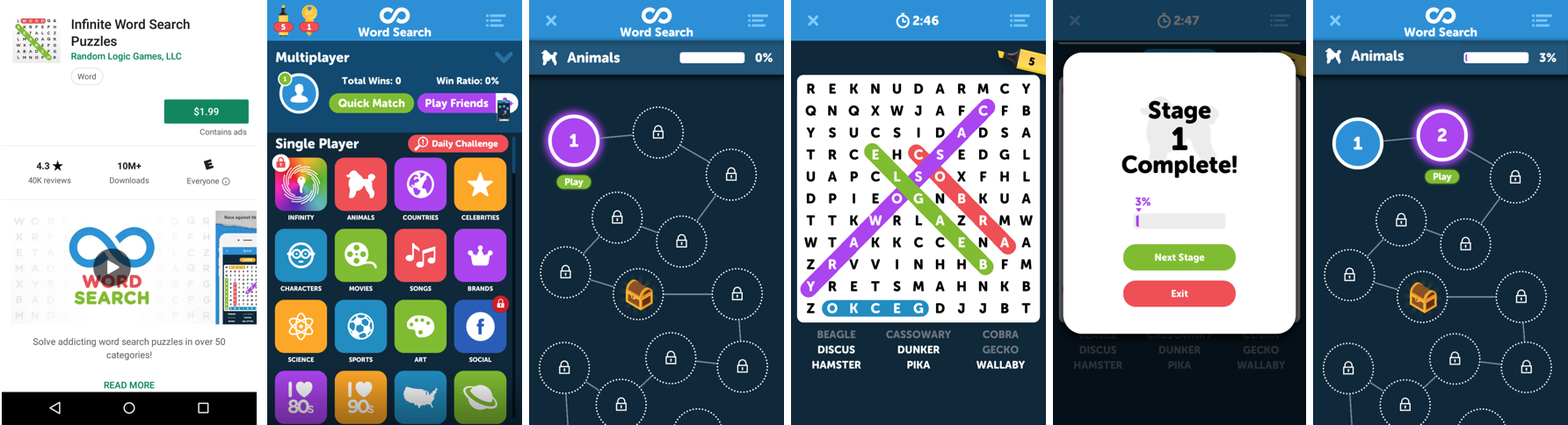}
\caption{PaidNoAds}\label{subfig:woody-PaidNoAds}
\end{subfigure}
\hspace{.04\textwidth}
\begin{subfigure}{.4\textwidth}
\includegraphics[width=\textwidth]{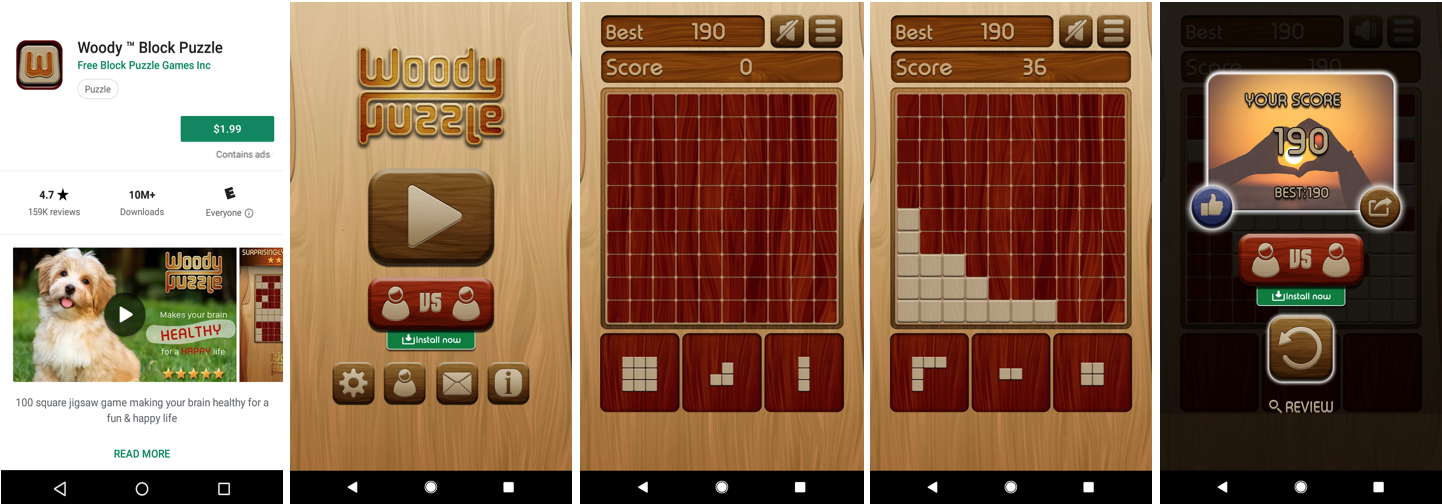}
\caption{PaidNoAds}\label{subfig:woody-PaidNoAds}
\end{subfigure}

\begin{subfigure}{.4\textwidth}
\includegraphics[width=\textwidth]{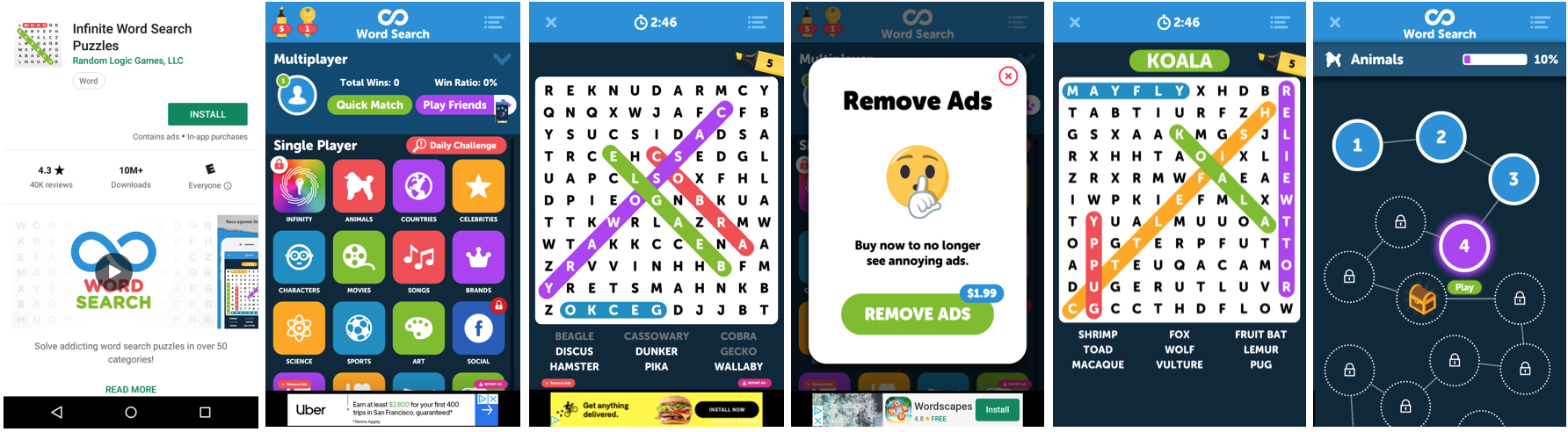}
\caption{BuyOutOfAds}\label{subfig:woody-BuyOutOfAds}
\end{subfigure}
\hspace{.04\textwidth}
\begin{subfigure}{.4\textwidth}
\includegraphics[width=\textwidth]{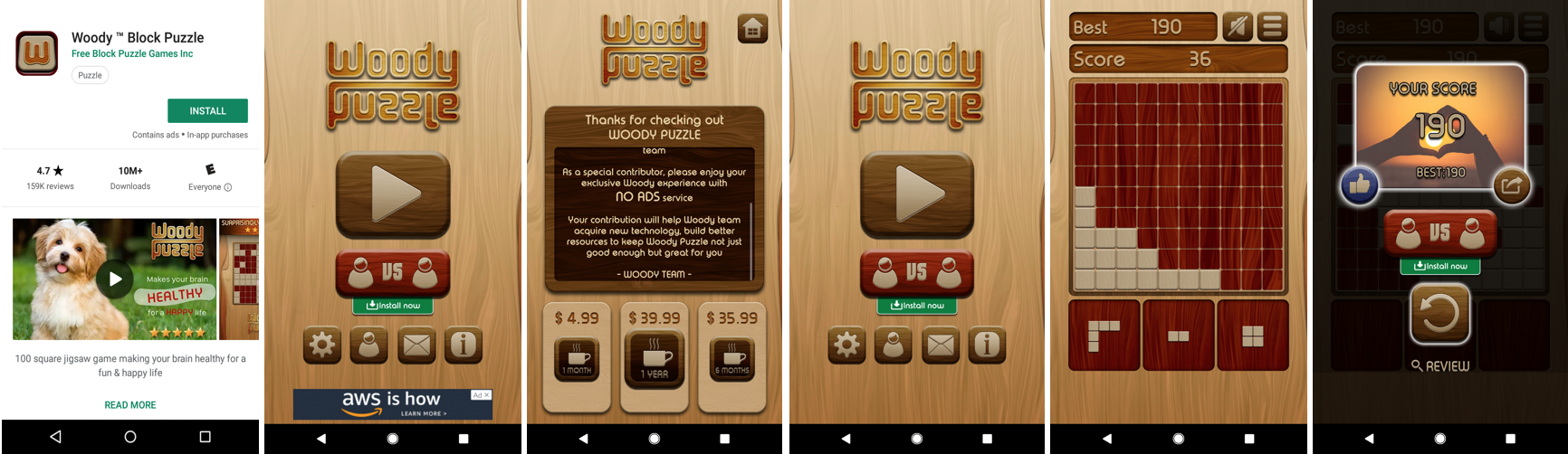}
\caption{BuyOutOfAds}\label{subfig:woody-BuyOutOfAds}
\end{subfigure}

\caption{The sequence of screenshots used in each condition. These screenshots were shown to the participant interwoven with a second-person story describing ``their'' interactions with the app. }\label{figure:summary-screenshots}
\end{center}
\end{figure*}

\subsection{Pilot Studies}
\label{subsec:pilots}

To begin, we ran two pilot studies ($n=250$ and $n=113$) to (1) identify what types of data people thought might be collected by mobile game apps and (2) to explore user beliefs about the purposes of data collection. We showed each participant a series of screenshots of a mobile game app (Woody Puzzle~\cite{WoodyPuzzle}, shown in Figure~\ref{figure:summary-screenshots}). In the first pilot, we then asked each participant what information they thought would be collected by the app, the company that created the app, or that company's business partners. In the second pilot, we then asked participants which of 19 data types (identified in the first pilot) they thought would be collected, and we asked open-ended questions about what they thought the purpose of the data collection was. Participants in each pilot received \$1.00 in compensation;the median completion time was 4.5 minutes for pilot 1 and 5.4 minutes for pilot 2. Two authors qualitatively coded each response. We identified 19 unique data types that users thought might be collected and 6 unique purposes (Figure~\ref{fig:pilotresults}). Any discrepancies or ambiguities were settled by discussing the example with a third author until a consensus was reached.  We used the results of these qualitative pilots in the subsequent full user study.

We chose to take this approach of generating choices through user responses to pilot studies because our hypotheses for the main study were focused on user perceptions and assessments of privacy threat. We believe that using categories of data defined by participants reflects users' actual beliefs about data collection better than, for example, using expert-generated categories of data with which participants in the main study might not be familiar, or which they might define differently than an expert.

\subsection{Full User Study}
\label{subsec:menthods-fullstudy}

In our full user study, we investigated three signals---presence of advertising, occurrence of payment, and timing of payment---to determine whether these signals affect user perceptions of the app.  We hypothesized that the presence of advertisements would cause users to perceive the app as posing a greater privacy threat---specifically, to believe that more data was collected and to trust the app less. This \emph{advertisement hypothesis} arose from past work which shows that the presence of ads does correlate with increased data collection~\cite{stevens2012, book2013longitudinal}\footnote{Although we note that the absence of visible ads does not necessarily denote the absence of ad libraries or the consequent effect on privacy, particularly for apps with both free and paid versions or apps where a user can pay to remove ads.} as well as our perception that advertisements are associated in the public consciousness with data collection. We also hypothesized that the occurrence of payment would cause participants to believe that the app poses less privacy threat. This \emph{payment hypothesis} arose from prior work that found that when apps offered both a free and a paid version, more data was collected by the free version~\cite{saborido2017comprehension}, as well as our perception that users might feel that it is possible to ``pay for privacy'', by making direct payments that enable the app to avoid a profit model involving ads or monetizing user data.

To test these hypotheses and explore the effect of these signals, we designed a study with five different conditions:
\begin{itemize}
    \item \textbf{FreeWithAds}: The user downloads a free app that contains ads. 
    \item \textbf{FreeNoAds}: The user downloads a free app that contains no ads.
    \item \textbf{PaidWithAds}: The user downloads a paid app that contains ads.
    \item \textbf{PaidNoAds}: The user downloads a paid app that contains no ads.
    \item \textbf{BuyOutOfAds}: The user makes an in-app purchase to remove ads from an otherwise free app. 
\end{itemize}
The first four conditions correspond to the different combinations of presence or absence of our two signals: ads and payment. The final condition is a variant of the \textbf{PaidNoAds} condition in which payment occurs as an in-app purchase after the app has been download instead of occurring when the app is initially installed, a payment model that is currently more common than \textbf{PaidNoAds} and which might be perceived differently by users. Our \emph{timing hypothesis} was that the timing at which a payment occurred would have no effect on users' assessment of the privacy risk posed by an app, that is that there would be no significant difference in assessed data collection or assessed trustworthiness between the \textbf{PaidNoAds} and the \textbf{BuyOutOfAds} conditions.

When selecting apps to use in our study, we elected to focus on game apps for two reasons: (1) because game apps---unlike other apps such as social media apps or banking apps---are unlikely to be perceived as inherent privacy threats simply due to the nature of the app and (2) because we believe that our experimental design had the most ecological validity in this domain, as both free and paid game apps with and without ads (and with options to buy out of ads) are notably present among the most popular apps on app stores. To enable us to evaluate whether users distinguish between apps with different data collection practices, we selected two different game apps from the list of most popular games on the Android Play Store in July 2019---Infinite Word Search or Woody Puzzle---and randomly assigned users to one of the two apps. 

Each participant was randomly assigned to one of the five conditions and shown a series of screenshots (Figure~\ref{figure:summary-screenshots}) accompanied with text describing a series of interactions with one of the two apps. These interactions and screenshots involved the presence of ads and payments as appropriate to the participant's condition.\footnote{Note that since neither app requested access to data controlled by just-in-time permissions, no such dialogues were shown in the story.}
 The participant was then asked to briefly summarize ``their'' interactions with the app in the story (our attention-check question). They were then asked which of the 19 data types they thought were collected by that app. For each type of data, the participant could select ``yes'', ``no'', or ``only in certain cases (e.g., if I give permission or if I click a button)''. For each type where the participant selected ``yes'', they were also asked the purpose of the collection. For each type of data we also asked the participant what signals caused them to mark that the data was collected, was sometimes collected, or was not collected. Finally, the participant was asked how trustworthy they considered the app using a 5-point Likert scale. The survey concluded with questions about technical background and demographics. The full survey is reproduced in Appendix~\ref{appendix:surveys}. Based on preliminary testing, we estimated that this survey would take participants 5 minutes to complete, so participants received \$1.00 compensation for completing the survey.

\paragraph{Analysis Plan.} Recognizing that the data types used in this survey might overlap, that not all data types  have equivalent privacy implications, and that the privacy impact of data collection might be nonlinear in the number of data types collected, we elected not to focus primarily on statistical tests that compared the mean number of data types collected. Instead, we used chi-squared tests to compare pairwise distributions of data between conditions, allowing us to test whether assessments of data collection and trust by users differed between conditions. For the purpose of these tests, we divided users into two buckets: those who believed that six or more different types of data were collected (\emph{high data collection assessment}) and those who believed that five or fewer different types of date were collected (\emph{low data collection assessment}). The dividing line of six data types was selected as this was the median number of data types our participants believed were collected. As this test is non-parametric, we do not need to meet any normality assumptions regarding the distribution of the data.

To further analyze the effect of various signals on data collection assessment and to validate the results of our chi-squared analysis, we fit a linear regression model to understand the predictive power of different factors on our data. This allowed us to see which predictors were important and how much of the variation in our results (i.e., effect size) were due to these predictors in the $R^2$ values.

For our analysis, we counted the number of data types that participants marked ``yes'' to (meaning that they believed the data type was collected by the app).  ``Only in certain cases (e.g., if I give permission or if I click a button)'' responses were categorized with the ``no'' responses (and analyzed as a no response throughout our paper) because our piloting suggested that participants did not perceive such data collection as a privacy threat.


\begin{figure*}
    \centering
    \begin{subfigure}[b]{.3\textwidth}
    \includegraphics[width=\textwidth]{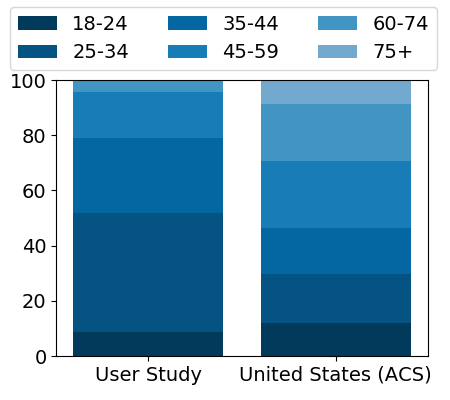}
    \caption{Age}
    \label{fig:demographics-age}
    \end{subfigure}
\centering
    \begin{subfigure}[b]{.3\textwidth}
    \includegraphics[width=\textwidth]{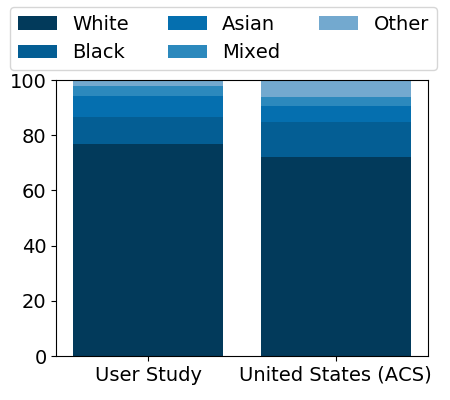}
    \caption{Race}
    \label{fig:demographics-race}
    \end{subfigure}
    \centering
    \begin{subfigure}[b]{.3\textwidth}
    \includegraphics[width=\textwidth]{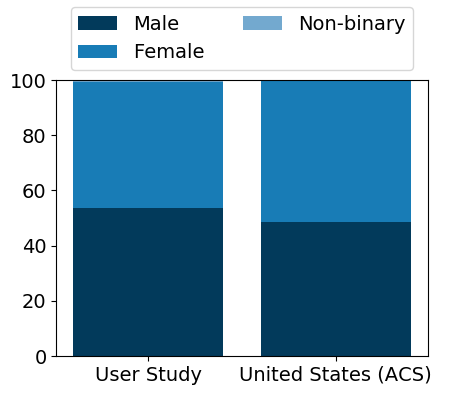}
    \caption{Gender}
    \label{fig:demographics-gender}
    \end{subfigure}
    \caption{Comparison between the demographics of our study participants and the demographics of the United States, as published in the American Community Survey (ACS)~\cite{ACS}.}
    \label{fig:demographics}
\end{figure*}

\subsection{Ethical Considerations}

We made consistent efforts throughout this work to minimize harms. Since this research is comprised of user studies that seek to understand user perceptions, the primary risk of this work is posed by the collection and handling of human subjects data. 

To minimize harm, we informed all potential study participants about the purpose of the study and about our data collection and use practices. No personally-identifiable information (e.g., names, email addresses, IP addresses) was collected. Users were also informed that they could opt out of the study at any time and any partial data about them would be deleted.

To fairly compensate users for their time, we performed preliminary pilot testing of each survey to estimate how long it would take for participants to complete and determined compensation based on a pro-rated wage of \$12 per hour using that estimate. We also tracked how long each participant spent on each survey to validate our compensation. For our pilot studies, the compensation was very close to our target of \$12 per hour (\$11.11-\$13.04). Unfortunately, we underestimated the completion time for our full study resulting in an average compensation of \$8.80. We subsequently modified the time estimation methodology for our lab to depend more strongly on estimates from individuals outside our research group, which improved our estimates. 
Since 2019, when these surveys were conducted, we have also increased our compensation target to track increases in minimum wage in U.S. states. 


All study materials and policies---including consent statements, survey instruments, participant compensation, and data collection practices (including data storage and dissemination)---were reviewed in advance by the Institutional Review Board (IRB) at our institution(s) and received and an IRB exemption approval.

\subsection{Limitations}
This study aimed to explore a small set of hypotheses relating to  advertisements, payments, and timing of payment. Our scope, therefore, limits us to findings related to these factors as well as findings which could be discovered by our experimental procedure. We expect that a variety of other factors may influence users' assessments of app data collection practices, and future work should explore such other factors. Similarly, our surveys used game apps, and results about mobile game apps may not generalize to other contexts, such as the web, or to other genres of application, such as health, banking, entertainment, and more. Future work should explore factors people use to assess data collection in other scenarios and types of apps.


Another potential limitation of our study is that it is unclear whether our results would generalize to types of data outside of the 19 discussed in our research. While we determined this list of 19 data types through extensive piloting---and therefore believe the 19 types of data used in our survey are comprehensive---it is unclear how robust these results are to changes in the list of data types considered. 

Finally, respondents to our pilot and full surveys were required to be located in the U.S. Since other cultures have different norms regarding data collection, it is unclear whether our results would remain the same with different survey populations. Future work should examine similar research questions in different cultural contexts.

\section{Results}
\label{sec:results}

We collected responses from 2109 participants; 145 were discarded for inconsistent or incoherent answers to the attention check question, and we analyzed the remaining 1964 responses. The demographics of our user study population, along with a comparison with the overall United States population, are summarized in Figure~\ref{fig:demographics}.

We observe briefly that the demographics of these user studies are not an exact match to the overall demographics of the United States; notably, our sample population is younger than the overall population, a known property of the Amazon Mechanical Turk worker pool~\cite{ipeirotis2010demographics}. However, prior work has found that Mechanical Turk results for surveys on security and privacy topics generalize well for most demographics despite this discrepancy~\cite{redmiles2019well}.

\begin{figure*}[t!]
\begin{subfigure}{.49\textwidth}
\begin{center}
      \includegraphics[width = \columnwidth] {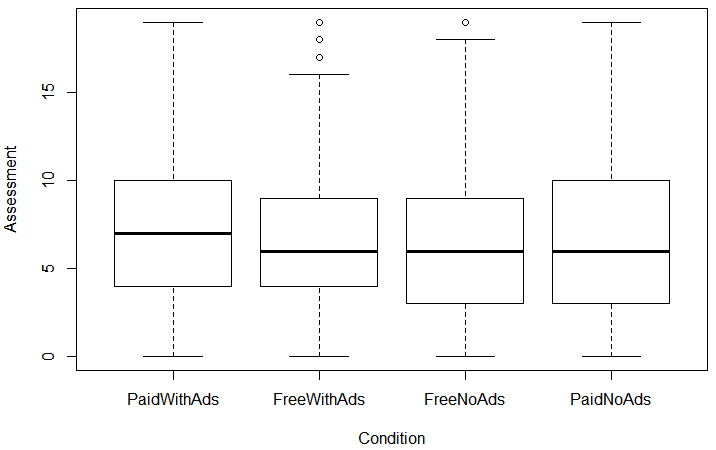}
      \caption{Boxplots showing of assessment of data collection (out of 19 possible types of data) by condition (without payment correction).}      \label{fig:assessmentNoCorrectionBoxplot}
\end{center}
\end{subfigure}
\hfill
\begin{subfigure}{.49\textwidth}
\begin{center}
      \includegraphics[width = \columnwidth] {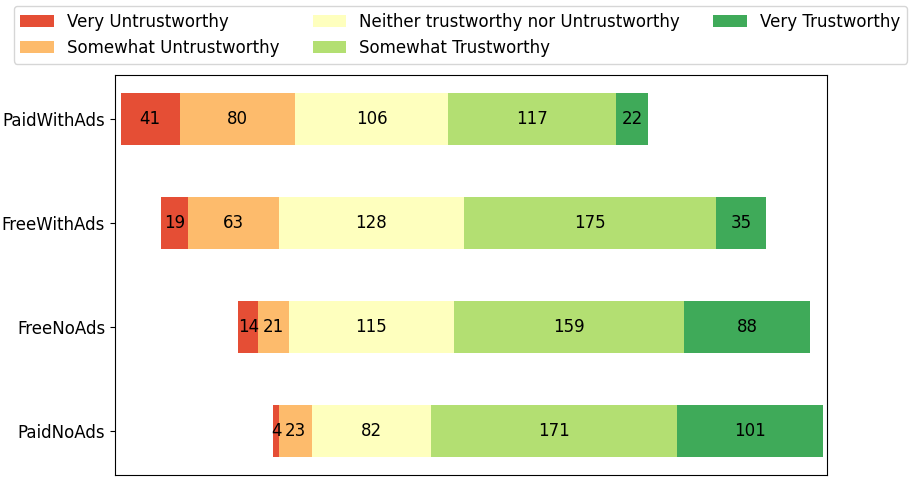}
      \hspace{12pt}
      \caption{User assessment of app trustworthiness.}      \label{fig:trustByCondition}
\end{center}
\end{subfigure}
\caption{The impact of advertising and payment on user assessment of privacy threat.}
\end{figure*}

\begin{table*}[t!]
\begin{center}
\begin{subtable}{.4\textwidth}
\begin{center}
\begin{tabular}{ l|r|c } 
\hline
Predictor & \multicolumn{1}{c|}{$p$-val} & Adjusted $R^2$\\ 
\hline
\hline
Ads & $0.001$ & $.00624$\\
Payment & $0.124$ & $.00087$\\
Ads \& Payment & $0.002$ & $.00731$\\
Trust & $<0.001$ & $.06200$\\
\hline
\end{tabular}
\end{center}
\caption{Linear regression analyses on users' assessment of data collection (without payment correction). }\label{table:regression-nocorrections}
\end{subtable}
\hspace{20pt}
\begin{subtable}{.4\textwidth}
\begin{center}
\begin{tabular}{ l|r|c } 
\hline
Predictor & \multicolumn{1}{c|}{$p$-val} & Adjusted $R^2$\\ 
\hline
\hline
Ads & $<0.001$ & $.0886$\\
Payment & $0.189$ & $.0005$\\
Ads \& Payment & $<0.001$ & $.1040$\\
\hline
\end{tabular}
\end{center}
\vspace{11pt}
\caption{\label{tab:trustAssessment} Linear regression analyses on users' trust. }
\end{subtable}
\end{center}
\caption{Linear regression analyses of users' assessment of privacy threat. Ads \& Payment denotes the regression where the predictor included an interaction between ads and payment.}
\end{table*}

All but 17 of our participants owned smartphones, suggesting that our online survey had successfully recruited smartphone users. We found that there were no statistically significant differences between users who had previously used the specific mobile app in the story and those who had not. We also found that there were no statistically significant based on respondents' self-reported level of technical background. 

\subsection{Advertising Hypothesis}

We analyzed our first four conditions (FreeWithAds and FreeNoAds, and PaidWithAds and PaidNoAds) to test our advertising hypothesis: that users interpret the presence of ads as a signal that an app poses a greater privacy threat. 

We found that the presence of advertising in mobile applications affected users' assessment of data collection: out of a total possible 19 pieces of data, users believed more data was collected on average in the conditions with ads (7.10 types of data) than the conditions without ads (6.36 types of data). A box plot summarizing these results for all four individual conditions is shown in Figure~\ref{fig:assessmentNoCorrectionBoxplot}. Using chi-squared tests, and we found that participants in conditions with advertising were significantly more likely to infer high levels\footnote{As defined in our analysis plan, assessment of six or more unique data types was classified as a high level of data collection.} of data collection ($p<.001$). Finally, we also ran linear regressions using presence of ads a predictor (Table~\ref{table:regression-nocorrections}); our results were consistent.

We also analyzed the effect of advertising on how trustworthy users consider an app to be. We found that users were significantly less likely to consider an app with ads to be somewhat or very trustworthy (44.4\% for conditions with ads compared to 66.7\% for the conditions without ads, $p<.001$). 
Users assessments of application trustworthiness for each condition are depicted in Figure~\ref{fig:trustByCondition}.  We also analyzed the effect of advertising on assessed trustworthiness using a linear regression (Table~\ref{tab:trustAssessment}); we found that the absence of advertising was a strong predictor of trust in an app, ($p < 0.001$) with an adjusted $R^2$ of $0.0886$. 

Together, these results constitute strong evidence to support our advertising hypothesis: users appear to interpret the presence of visible ads as a signal that an app poses a privacy threat.

\paragraph{Evaluating Ads as a Data Collection Signal.} There is some evidence that the presence of ads does correlate with increased data collection~\cite{stevens2012, book2013longitudinal}.  This means that users' reliance on ads as a proxy for data collection is, in some sense, reasonable and warranted. Users might thus be making more informed decisions by using signals to assess data collection that actually correlate with data collection. 
However, we found that that was not the case: individual accuracy at assessing which data types were collected by our two apps (Woody Puzzle and Infinite Word search) was low, and there was no significant difference in assessment between the two apps.

We performed expert readings of the privacy policies of both apps to establish a ground-truth of the data collected by each app. Additionally, we read the privacy policies of ad libraries embedded by the apps. We broke down the data collected by an app into two categories: data collected directly by the app, and data collected by ad libraries embedded in the app. The data collected directly by the app was considered to be the ground truth of how much data would be collected for conditions without ads.\footnote{We note that there is no guarantee that an absence of visible ads necessarily corresponds to decreased data collection by a particular app; apps that offer free (with ads) and paid (ad-free) versions sometimes contain the same ad libraries~\cite{han2020,bamberger2020}---with the corresponding privacy implications---and advertising-free apps might in some cases collect more data for other purposes (e.g., to aggregate or sell). Nonetheless, we view these definitions as a reasonable approximation for data collection in the various different conditions.} We took the union of an app's individual collection with its ad libraries' collection to determine the ground truth of how much data would be collected in conditions with ads. We found that Woody Puzzle collected 8 (7 omitting collection by ad libraries) types of data, and that Infinite Word Search collected 7 (3 without ad libraries) types of data.
    
Overall, our respondents had an accuracy of $67.4\%$ in predicting which data types were collected by the app they were shown. However, this number is influenced by the number of data types that were (correctly predicted to) not be collected by these apps. If we only consider data types collected by the app, users correctly identified that those data types were collected only 55.1\% of the time, a rate that is not much higher than guessing at random. Moreover, there was no significant difference in users' assessment of data collection between the two apps ($p=0.166$), which highlights the fact that correlations inherently fail to capture the particular data collection practices of individual apps, particularly those that deviate from the norm.  
    
Based on these results, we believe that use of proxy signals---even ones that strongly correlate with actual data collection---should not be considered a sufficient basis for informed consent, or as evidence that the app ecosystem is behaving appropriately with respect to privacy. While these heuristics lean the right way, they rarely match the specifics of data collection practices. These specifics are often extremely important to users, especially for users for whom privacy is a matter of safety. For example, specific types of data, such as location data or contact information, may be highly sensitive for specific groups, such as survivors of intimate partner violence~\cite{freed2018stalker}. Instead, these correlations should be taken as evidence of social norms around data collection---specifically that apps without ads  should not collect additional personal information beyond that required for app functionality.  
    
Additionally, participants were unable to accurately predict differences in data collection practices between the two game apps in our study. Our results are compatible, therefore, with the idea that users are ``stereotyping'' apps using clues such as the presence of ads and, perhaps, the genre of the app. Apps that choose to collect less data are unlikely to be rewarded with a competitive advantage. 
Instead, these results suggest that we need revised app store guidelines and rules to enforce compliance with social norms around data collection by mobile apps.

\subsection{Payment Hypothesis}

Next, we analyzed our first four conditions (FreeWithAds and PaidWithAds, and FreeNoAds and PaidNoAds) to test our payment hypothesis: that user interpret the absence of payment as a signal that an app poses a greater privacy threat (i.e., collects more data and/or is less trustworthy).

\begin{figure}[t!]
      \includegraphics[width = \columnwidth] {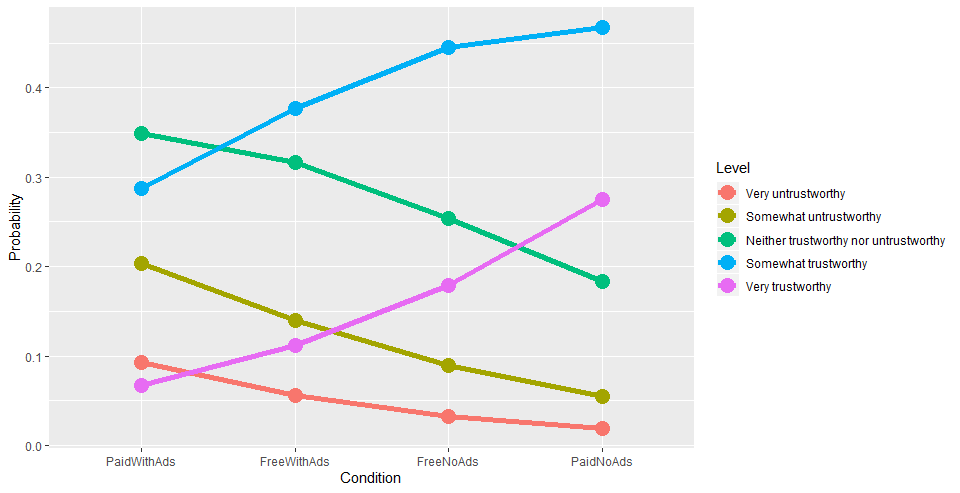}
      \caption{Predicted probabilities from an ordinal logistical regression model for how trustworthy users would find an app within each condition.}
      \label{fig:trustOrdinalRegr}
\end{figure}

We started by evaluating the effect of payment on users' perceptions of data collection. To our surprise, we found that occurrence of payment actually caused higher data collection assessment (on average, 6.91 types of data of paid apps compared to 6.56 types for free apps), although a chi-squared test found that this difference was not statistically significant ($p=.169$). Our linear regressions using presence of ads and payment as predictors returned consistent results (Table~\ref{table:regression-nocorrections}); when considering an interaction between ads and payment, the predictor ($p = .002$) explained more of the variation in data collection assessment, with an adjusted $R^2$ of $.007$, but payment alone was not a significant predictor.  These results provide insight into earlier findings that payment and advertising in combination affect user perceptions of apps (i.e., that users expect a paid, ad-free version of the app will have better security and privacy behaviors than a free, ad-supported app~\cite{han2020,bamberger2020}). They suggest that when these signals occur in combination---that is, when a user is actually offered a choice between a free, ad-supported app and a paid, ad-free app---users believe that they are indirectly buying privacy by installing an ad-free version of an app rather than believing that they are directly paying for privacy.  

We also evaluated the effect of payment on users' perception of app trustworthiness. We found that payment alone did not result in significant differences is assessed trustworthiness; instead, there appears to be an interaction between trustworthiness and payment. Users were most likely to consider paid apps with no ads to be somewhat or very trustworthy (71.5\%) whereas users were \emph{least} likely to consider paid apps with visible ads to be somewhat or very trustworthy (37.9\%). The free conditions without and with ads feel somewhere in the middle (50.0\% and 62.2\%, respectively). Chi-squared tests found that all pairwise differences between conditions were statistically significant ($p<.01$). Users assessments of application trustworthiness for each condition are depicted in Figure~\ref{fig:trustByCondition}.  

\begin{table*}
\begin{center}
\begin{tabular}{ l|c||l|c||l|c||l|c } 
\hline
Data Type & \% & Data Type & \% & Data Type & \% & Data Type & \%\\ 
\hline
credit card info & $68.9$ & phone number & $9.5$ & interests & $2.4$ & inter. w/ ads  & $1.6$\\
mailing addr & $17.4$ & photos & $4.0$ & phone identifier & $2.3$ & inter. w/ app & $1.4$\\
email & $12.2$ & A/V sensor data & $3.6$ & basic demo. & $2.2$ &   inter. w/ other apps  & $1.4$\\
phone account info & $11.4$ & contact list & $3.5$ & IP address & $2.0$ & general phone info & $1.3$ \\
unique identifiers & $10.4$ & sensitive demo.  & $3.4$ & current location & $1.9$& & \\
\hline
\end{tabular}
\end{center}
\caption{\label{tab:paymentMisconception} Percent users who believe various data types are collected for the purpose of processing payments. }
\end{table*}

Given this interaction, we also analyzed the effect on assessed trustworthiness with a linear regression. 
As in predicting data collection assessment, the presence of payment alone was not a good predictor of perceived trustworthiness of an app ($p = 0.189$) but rather amplified the effects of advertising. Thus, the condition to which a user was randomly assigned was slightly stronger as a predictor than the presence of advertising alone ($p < 0.001$, adjusted $R^2 = 0.104$). The extra predictive power comes from the interaction of payment with advertising. These results are summarized in Table~\ref{tab:trustAssessment}. We further explored the interaction between ads and payment by running an ordinal logistic regression to evaluate how well the combination of these signals predicted user assessments of trustworthiness. We again found that payment amplifies the effect of ads on trustworthiness. As shown in Figure~\ref{fig:trustOrdinalRegr}, the likelihood of a user finding the app ``somewhat'' or ``very trustworthy'' is highest in the condition with payment and no ads, and lowest in the condition with both payment and ads. The reverse is true for untrustworthy assessments. 
We also found that how trustworthy a user rated the app was a strong predictor of data collection assessment ($p < 0.001$,) with an adjusted $R^2$ of $0.062$ on a linear regression model; these results are summarized in Table~\ref{table:regression-nocorrections}.



\paragraph{A Mental Model for Payment Amplification.} Our findings of a non-linear amplification effect of payment surprised us by refuting our simpler hypothesis that occurrence of payment, like absence of advertisement, would be associated directly with the perception of reduced data collection. While our results cannot fully explain the causes of this effect, we present here one model we believe to be compatible with our results. 
    
We hypothesize that users believe apps profit in one of two ways: either apps monetize user data, or they charge users directly. Under this ``buying privacy'' model, users would believe that payment indicates lower levels of data collection, since direct payment negates the needs to make money by monetizing user data. We hypothesize that this signal may in fact be present in the data, showing itself in the fact that for apps with no ads, the presence of payment was associated with a belief that less data was collected. 
    
Further, we hypothesize that this signal may be covered up when ads are present in the app, because the combination of ads and payment drive trust low enough that the effects of trust dominate over the effects of the ``buying privacy'' model alone. 
Our results show that apps with both ads and payment were considered significantly less trustworthy.
In particular, significantly more users rated the conditions with ads to be very or somewhat untrustworthy (Figure~\ref{fig:trustByCondition}), levels where trust causes very large increases in assessment of data collected. The buying privacy effect may be present here, but it is dominated by the effects of trust. Meanwhile, on the high end, variations in trust have no effect on data collection assessment, since we found no difference in assessment when apps were considered neutrally trustworthy or better. Thus in this situation, the ``buying privacy'' theory shines through and explains why users believed that even less data was collected in scenarios with both payment and ads.

    
This model is one possible explanation. While compatible with our results, it is best viewed as a hypothesis to be further explored by future work. We believe that exploring people's beliefs and mental models about the profit models of apps, and the inferences they make about data collection based on those mental models, could be a valuable future direction toward understanding the full suite of features that people use to assess the privacy properties of the apps they use. Such modeling could be invaluable to evaluating the ways in which informed consent may be failing to uphold a  well-functioning privacy ecosystem.

\paragraph{Misconceptions about Payments.} Our result that found that users consistently associated occurrence of payment with a (slight) increase in data collection was inconsistent with our initial payment hypothesis: that users would associate payment with a decrease in data collected (i.e., that payment would be assumed to ``buy privacy''). This surprising finding caused us to take a closer look at the data collected in our user study; on closer inspection, we discovered widely held misconceptions about how payments are processed and about what information is collected by an app in order to process payments.  Overall, 24.7\% of our users believed that one or more data type was collected by the app for the purpose of payment processing. In actuality, no data is collected by the app for the purpose of processing payments; all payments processing for app purchases and in-app purchases is performed by the app store, and payment-related data is not collected by or shared with the app. Unsurprisingly, we observed that this misconception was most common in the conditions in which payment occurred (28.1\% of PaidWithAds, 34.6\% of PaidNoAds, 37\% BuyOutAds) and that this misconception was most common for data types associated with credit card payments (e.g., credit card info and mailing address). Complete information about frequency of payment misconceptions by data type is given in  Table~\ref{tab:paymentMisconception}.

Since data collection for the purpose of processing payments can be considered contextually appropriate and therefore not a violation of user privacy~\cite{nissenbaum2004privacy}, we re-analyzed our four main conditions with a payments correction. Specifically, we treated any data type the user believed to be collected for the purpose of processing payments as though they had reported that data type would not be collected. We believe this more accurately reflects user perceptions about the privacy threat posed by a mobile app.

\begin{figure}[t!]
      \includegraphics[width = \columnwidth] {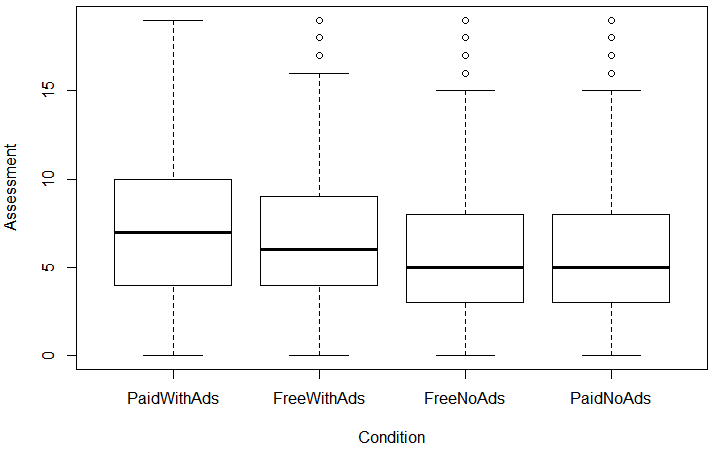}
      
\caption{Boxplots showing of assessment of data collection (out of 19 possible types of data) by condition (with payment correction).}\label{fig:assessmentCorrectedBoxplot}
\end{figure}

After these corrections, the mean number of data types believed to be collected in each condition was as follows: PaidWithAds (7.011), FreeWithAds (6.636), FreeNoAds (6.139), PaidNoAds (5.858). A  box plot is shown in Figure~\ref{fig:assessmentCorrectedBoxplot}.

After applying this correction, we found that advertising is still a statistically significant signal (with an even lower $p$-value), and that payment is still not ($p=.56$). However, a clearer pattern emerged when we ran the same linear regressions on the corrected data. The presence of advertising as an indicator variable was a stronger predictor of assessment of data collection ($p < 0.001$) with an adjusted $R^2$ of $0.0082$. Moreover, while the presence of payment was once again not significant in predicting the amount of data users believed was collected, payment as a signal \emph{amplified} the effects of advertising. That is, users believe that more data is being collected when payment occurs and advertising is present, and they believe that less data is being collected when payment occurs and there are no ads. These two variables in combination (i.e., with interaction) explain more of users' assessment of data collection ($p = .001$) with an adjusted $R^2$ of $0.0084$.

\begin{table}[t!]
\begin{center}
\begin{tabular}{ l|r|c } 
\hline
Predictor & \multicolumn{1}{c|}{$p$-val} & Adjusted $R^2$\\ 
\hline
\hline
Ads & $<0.001$ & $.00820$\\
Payment & $0.895$ & $-.00063$\\
Ads \& Payment & $0.001$ & $.00841$\\
Trust & $<0.001$ & $.06990$\\
\hline
\end{tabular}
\end{center}
\caption{Linear regression analyses on users' assessment of data collection (with payment correction). Ads \& Payment is the regression where the predictor contained the interaction of the two terms.}\label{table:regression-corrections}
\end{table}

With this correction, trustworthiness remained the strongest predictor of users' assessment of data collection ($p < 0.001$, adjusted $R^2$ $0.07$) on a linear regression model, and thus with more predictive power than without corrections for payment misconceptions. These results are summarized in Table~\ref{table:regression-corrections}.

\subsection{Timing Hypothesis}
Two of our conditions, BuyOutOfAds and PaidNoAds, involved the user making a payment and then receiving an ad-free game experience. However, the timing of the payment differed between these two conditions: in the PaidNoAds condition the user paid to download an app with no ads, whereas in BuyOutOfAds condition the user downloaded an app with ads and then made an an in-app purchase to remove the ads. Since users in the BuyOutOfAds condition were instructed to consider only data collection that occurred after they had pay to remove ads, we hypothesized that the difference in the timing of the payment would not significantly affect respondents' data collection and trust selections. 

\begin{figure}[t!]
\begin{center}
      \includegraphics[width = \columnwidth] {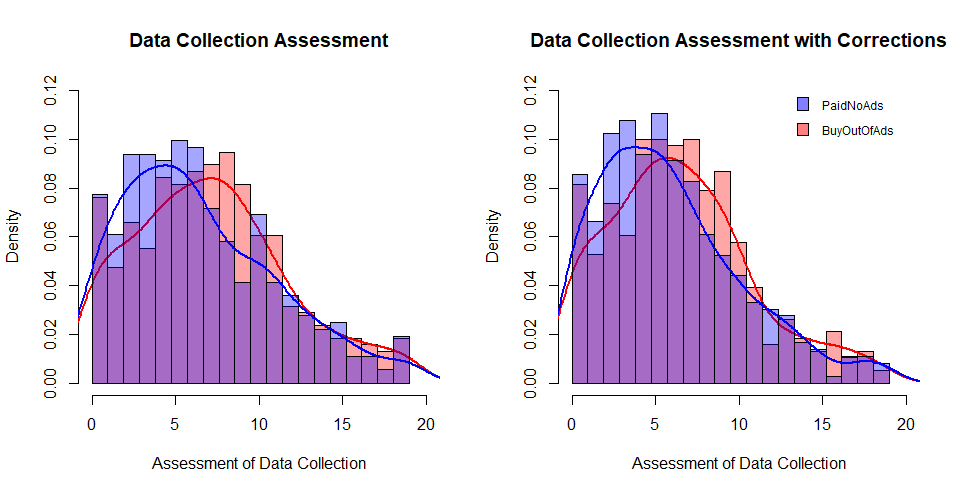}
      \caption{Comparison of users' assessment of data collection between the PaidNoAds and BuyOutAds conditions. The bars are the density distributions and the lines represent the density curves for each condition.}      
      \label{fig:pairwiseDensity}
\end{center}
\end{figure}
\begin{figure}[t!]
\begin{center}
      \includegraphics[width = \columnwidth] {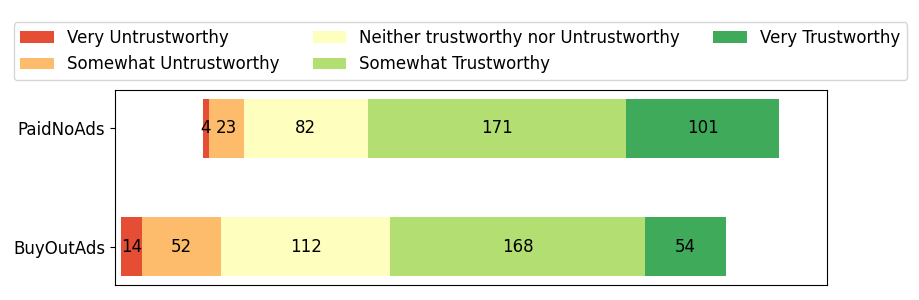}
      \caption{Comparison of users' assessment of app trustworthiness between the PaidNoAds and BuyOutAds conditions.}      \label{fig:trustBuyOut}
\end{center}
\end{figure}

However, our results disproved our timing hypothesis: we instead found that user assessment of these two conditions was significantly different. Users in the BuyOutOfAds condition believed that significantly more data was collected by the app (after the purchase occurred) compared to users assigned to the PaidNoAds condition ($p=.003$). This difference persisted even after applying the payment correction ($p=.017$), which suggests that the timing of payment has a significant impact on user beliefs about data collection that is unrelated to payments processing. With the payment correction applied, 46.25\% of respondents in the BuyOutOfAds condition inferred that high levels of data collection occurred, whereas only 37.53\% of respondents in the PaidNoAds condition inferred high levels of data collection.\footnote{As described in Section~\ref{subsec:menthods-fullstudy}, an inference that six or more unique data types were collected was classified as a high level of data collection.}  A more detailed comparison between the data collection assessments for these two conditions is depicted in Figure~\ref{fig:pairwiseDensity}; the fact that data collection assessments are shifted left for the BuyOutOfAds compared to the PaidNoAds condition shows that participants assess higher levels of data collection---and thus higher privacy threat---in the BuyOutOfAds condition compared to the PaidNoAds condition.

We also evaluated whether the timing of payment affected how trustworthy users considered an app. We found that users in the BuyOutOfAds condition reported significantly lower levels of trust than users in the PaidNoAds condition ($p<.001$). In the BuyOutOfAds condition, 55.5\% of respondents reported that they considered the app to be ``very trustworthy'' or ``somewhat trustworthy'', whereas 71.39\% of users considered the app in the PaidNoAds condition to be trustworthy. These responses are summarized in Figure~\ref{fig:trustBuyOut}.   This indicates that when payment occurs as an in-app purchase after the app has been downloaded, users are less likely to trust the app. 

Given that app developers have increasingly moved away from offering two versions of their app---an ad-supported free version and a paid version with no ads---in favor of a single version of the app with an option to remove ads as an in-app purchase, this result might have significant implications for overall trust in mobile applications.

\subsection{Limitations}

A key limitation of this work is that it relies on a simulation study: participants were asked to imagine a series of interactions with an app. This study design was adopted due to practical feasibility constraints. Since the vast majority of our participants were smartphone owners, we believe that our participants were able to accurately identify their assessments. Nonetheless, confidence in these results would be enhanced by replication in a study using an actual app.

Both for practical reasons and to avoid confounding effects of inherent privacy concerns about certain classes of mobile apps, our study focused exclusively on mobile game apps. Results on mobile game apps might not generalize to other contexts, such as the web, or to other genres of application, such as health and banking, and productivity; future work should explore signals that impact evaluations of privacy threat in other domains. 



Finally, respondents to our pilot and full surveys were required to be located in the U.S. Since other cultures have different norms regarding data collection, it's unclear whether our results would remain the same with different survey populations. Future work should examine similar research questions in different cultural contexts.

\section{Conclusion}
\label{sec:conclusion}

This work was motivated by the question: if privacy policies do not effectively inform users about data collection practices, how do users assess the privacy threat from mobile apps? Our results (1) demonstrate that users rely  on the presence or absence of advertising, (2) identify a non-linear, amplifying effect of payment, and (3) identify how timing of payment and misconceptions about payment affect user beliefs about data collection and trustworthiness. 
However, we also find that users are not able to correctly identify the types of data that are collected, and that users do not distinguish between apps with different data practices.

These results provide insight into how user-visible signals impact perceptions of privacy threat; they also give insight into social norms around data collection by mobile apps, specifically that apps without ads (and especially paid apps without ads) should not collect additional information beyond that necessary for app functionality but that additional data collection by apps with ads is expected (e.g., for the purposes of improving targeted ads).
These results suggest that the using these signals as proxies for data practices is insufficient to establish informed consent. They also suggest that privacy would be enhanced if app stores modified their rules and guidelines to enforce that apps comply with social norms about data collection~\cite{Niss09,Niss11}. 

Further work will be required to definitively establish how these signals correlate with actual data collection practices, to determine whether there exist other factors that influence user assessment, and to extend the scope beyond game apps. Nonetheless, we view this work as an important step in understanding how users assess the privacy of mobile apps, what factors influence those evaluations, and how those factors intersect with social norms.


\bibliographystyle{plain}
\bibliography{_main.bib}

\appendix
\section{Survey Questions}\label{appendix:surveys}

This Appendix contains the list of questions asked during our full user study. The possible answers to questions 3,4,5,6, and 7 were determined based on qualitative coding of the answers to open text-entry questions posed in the pilot studies.  We also asked participants demographic questions.

We recruited participants on Amazon Mechanical Turk; the survey was advertised as ``A short survey to understand ordinary people’s perceptions of privacy in mobile apps''. Respondents were paid $\$1.00$ for completing the survey, which took an average of 6.8 minutes. Respondents were required to be located in the U.S. and to have previously completed at least 50 HITs with an approval rate of at least 95\%.

The free-response question (asking the users to summarize the story) was used as an attention check question: responses that were nonsensical, irrelevant, or did not discuss the app or the story were rejected (145 responses). The remaining 1964 respondents became our study population.

The survey questions are as follows:\\

\begin{enumerate}
    \item Please briefly summarize the above story. \\
    \indent \textit{For this question, the user is given a text box to write a free-response answer in.}\\
    
    \item How trustworthy do you consider this app?
    \indent \textit{This question is either shown to the user in this order or reversed order.}
    \begin{itemize}
        \item Very trustworthy
        \item Somewhat trustworthy
        \item Neither trustworthy nor untrustworthy
        \item Somewhat untrustworthy
        \item Very untrustworthy\\
    \end{itemize}
    
    \item For each piece of information listed below, do you think this app collects this information (after you have removed the ads)?\\
    \indent \textit{The question contains the text in parentheses if the condition is BuyOutOfAds. For each piece of information, the user can choose ``Yes," ``Only in certain cases (e.g., if I give permission, or if I click on a button)," or ``No." The following pieces of information are presented to each user in a random order.}\\
    \begin{itemize}
        \item phone number
        \item email
        \item phone account information (e.g., iCloud account or Google account username)
        \item physical mailing address
        \item credit card information
        \item information that uniquely identifies the user (e.g., your name)
        \item phone identifier (i.e., serial number)
        \item general information about your phone (e.g., model, screen size, OS, carrier)
        \item current IP address
        \item basic demographics (e.g., age)
        \item sensitive demographics (e.g., religion, race, sexual orientation)
        \item current location (e.g., GPS location or WiFi networks you're connected to)
        \item interactions with the app (e.g. what levels the user reaches)
        \item interactions with ads in the app (e.g., clicks)
        \item interactions with other apps
        \item sensor data from the camera and/or microphone
        \item photos
        \item contact list
        \item information about the user's interests\\
    \end{itemize}
    
    \item You said that each of the following types of information is collected by the app. What is the primary reason that caused you to believe this? \\
    \indent \textit{For this question, all the pieces of data for which the user responded ``Yes" in question 3 are listed. For each piece of data, the user must choose between the following options.}\\
    \begin{itemize}
        \item Most apps collect this
        \item Most game apps collect this
        \item Payment option
        \begin{itemize}
            \item Because the app was free - \textit{FreeWithAds and FreeNoAds conditions}
            \item Because your interactions with the app included payment - \textit{BuyOutOfAds condition}
            \item Because you paid for the app - \textit{PaidWithAds and PaidNoAds conditions}
        \end{itemize}
        \item Because of the buttons in the app (e.g., multi-player, share, like)
        \item Because of the app's overall design, layout, or quality
        \item Ads option
        \begin{itemize}
            \item Because the app displays ads - \textit{FreeWithAds and PaidWithAds conditions}
            \item Because the app doesn't display ads - \textit{FreeNoAds and PaidNoAds conditions}
            \item Because the app no longer displays ads - \textit{BuyOutOfAds condition} \\
        \end{itemize}
    \end{itemize}
    
    \item You said that each of the following types of information are collected by the app only in certain conditions (e.g., when you give permission or when you click on a certain button).  What is the primary reason that caused you to believe that this information won't be collected otherwise?\\
    \indent \textit{For this question, all the pieces of data for which the user responded ``Only in certain cases (e.g., if I give permission, or if I click on a button)" in question 3 are listed. For each piece of data, the user must choose between the following options.}\\
        \begin{itemize}
        \item Most apps only collect this in certain circumstances
        \item Most game apps only collect this in certain circumstances
        \item Payment option
        \begin{itemize}
            \item Because the app was free - \textit{FreeWithAds and FreeNoAds conditions}
            \item Because your interactions with the app included payment - \textit{BuyOutOfAds condition}
            \item Because you paid for the app - \textit{PaidWithAds and PaidNoAds conditions}
        \end{itemize}
        \item Because of the buttons in the app (e.g., multi-player, share, like)
        \item Because of the app's overall design, layout, or quality
        \item Ads option
        \begin{itemize}
            \item Because the app displays ads - \textit{FreeWithAds and PaidWithAds conditions}
            \item Because the app doesn't display ads - \textit{FreeNoAds and PaidNoAds conditions}
            \item Because the app no longer displays ads - \textit{BuyOutOfAds condition} \\
        \end{itemize}
    \end{itemize}
    
    \item You said that the app does not collect the following pieces of information. What is the primary reason that caused you to believe that the app doesn't collect each piece of information?\\
    \indent \textit{For this question, all the pieces of data for which the user responded ``No" in question 3 are listed. For each piece of data, the user must choose between the following options.}\\
    \begin{itemize}
        \item Because the app has no way to collect this
        \item Most apps don't collect this
        \item Most game apps don't collect this
        \item Payment option
         \begin{itemize}
            \item Because the app was free - \textit{FreeWithAds and FreeNoAds conditions}
            \item Because your interactions with the app included payment - \textit{BuyOutOfAds condition}
            \item Because you paid for the app - \textit{PaidWithAds and PaidNoAds conditions}
        \end{itemize}
        \item Because of the buttons in the app (e.g., multi-player, share, like)
        \item Because of the app's overall design, layout, or quality
        \item Ads option
        \begin{itemize}
            \item Because the app displays ads - \textit{FreeWithAds and PaidWithAds conditions}
            \item Because the app doesn't display ads - \textit{FreeNoAds and PaidNoAds conditions}
            \item Because the app no longer displays ads - \textit{BuyOutOfAds condition} \\
        \end{itemize}
    \end{itemize}
    
    \item In your opinion, for what purpose would this app collect each of these pieces of information (after you have removed the ads)? Please select the option that best describes your opinion.\\
    \indent \textit{The question contains the text in parentheses if the condition is BuyOutOfAds. For this question, all the pieces of data for which the user responded ``Yes" in question 3 are listed. For each piece of data, the user must choose between the following options.}\\
    \begin{itemize}
        \item To sell to other companies
        \item To make more money from ads
        \item To identify who uses their app
        \item To process payments
        \item To perform necessary app functionality
        \item To improve the app (e.g. better performance, design/layout)
        \item There is no particular purpose\\
    \end{itemize}

\end{enumerate}

\end{document}